\documentclass[runningheads]{llncs}
\usepackage[utf8]{inputenc}
\usepackage{authblk}
\usepackage{setspace}
\usepackage[margin=1.25in]{geometry}
\usepackage{graphicx}
\graphicspath{ {./figures/} }
\usepackage{subcaption}
\usepackage{amsmath}
\usepackage{tabularx,ragged2e, booktabs}
\usepackage{hyperref}
\usepackage[utf8]{inputenc}
\usepackage{multirow}
\usepackage{enumitem}
\usepackage{booktabs}
\usepackage{tikz}
\usetikzlibrary{matrix}
\usetikzlibrary{matrix, positioning}

\newcolumntype{Y}{>{\raggedright\arraybackslash}X}

\usepackage{natbib}
\title{Fair and Inclusive Participatory Budgeting: Voter Experience with Cumulative and Quadratic Voting Interfaces}

\author{Thomas Wellings\inst{1} \and
Fatemeh Banaie Heravan\inst{1} \and
Abhinav Sharma\inst{1} \and
Lodewijk Gelauff\inst{2}, \\
Regula Hänggli Fricker\inst{3} \and
Evangelos Pournaras\inst{1}
}

\authorrunning{Wellings et al.}

\institute{
  University of Leeds, United Kingdom {t.wellings, f.banaieheravan, a.sharma1, e.pournaras\}@leeds.ac.uk} \and
  Stanford University, United States of America {lodewijk@stanford.edu} \and
  University of Fribourg, Switzerland {regula.haenggli@unifr.ch}
}

\begin{document}

\maketitle

%%%%%% Abstract %%%%%%
\begin{abstract}

Cumulative and quadratic voting are two distributional voting methods that are expressive, promoting fairness and inclusion, particularly in the realm of participatory budgeting. Despite these benefits, graphical voter interfaces for cumulative and quadratic voting are complex to implement and use effectively. As a result, such methods have not seen yet widespread adoption on digital voting platforms. This paper addresses the challenge by introducing an implementation and evaluation of cumulative and quadratic voting within a state-of-the-art voting platform: Stanford Participatory Budgeting. The findings of the study show that while voters prefer simple methods, the more expressive (and complex) cumulative voting becomes the preferred one compared to k-ranking voting that is simpler but less expressive. The implemented voting interface elements are found useful and support the observed voters' preferences for more expressive voting methods. 
 
\end{abstract}

%%%%%% Main Text %%%%%%

\section{Introduction}

Digital voting platforms have granted the opportunity for increased citizen participation and inclusive decision making~\citep{Pournaras2020,Helbing2023,junior2022informative}. For instance, in participatory budgeting community members directly decide how to spend a public budget, fostering a bottom-up decision-making process that encourages civic engagement and transparency which can improve quality of life~\citep{wellings2023improving}. However, optimizing online voter experience is challenging, especially when implementing complex voting methods to use in participatory budgeting campaigns, for instance, cumulative and quadratic voting. These methods require a user interface (UI) with which voters can convey complex preferences, but without sacrificing simplicity to maintain an engaging voter experience. Similarly, it was initially argued that the use of knapsack voting would be too burdensome~\citep{benade2017preference} as voters need to approve projects that do not surpass the available budget. However, it was later demonstrated that it was possible to implement knapsack voting with the appropriate voter interface~\citep{goel2019knapsack}. Therefore, in this paper, we conduct an empirical study to assess and optimize voter experience with cumulative and quadratic voting on the Stanford Participatory Budgeting platform. This study provides new insights about the integration of complex voting methods to digital voting platforms.

\begin{figure}
  \centering
  \includegraphics[width=0.8\textwidth]{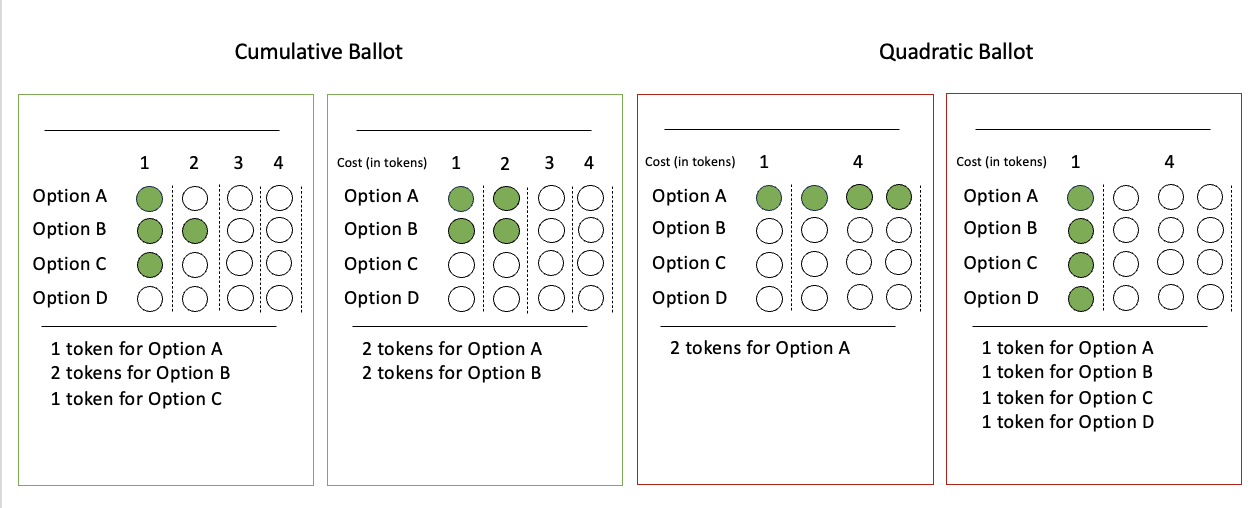}
  \caption{Example of cumulative and quadratic ballots. The two ballots to the left highlight an example of cumulative voting. The two ballots on the right highlight an example of quadratic voting. Notice that is not possible to distribute more than one token for two options in quadratic voting, due to the cost of tokens increasing quadratically.}
  \label{fig:com}
\end{figure}

Cumulative voting~\citep{kato2021positionality} and quadratic voting~\citep{lalley2014nash} are part of a group of voting methods that allow voters to distribute their votes across multiple candidates in elections or projects in participatory budgeting. It could lead to more inclusive and representative results~\citep{ricnckevivcs2013equality}, increasing minority's turnout relative to the majority, and the minority's share of winning seats/projects~\citep{casella2023minority}. Specifically, cumulative voting enables voters to express their preferences in a non-linear way (i.e., it means that they can distribute a fixed number of points across different options). This approach is different from ranked voting, where the voter ranks the options, or approval voting, where voters pick up the options they prefer. Quadratic voting is similar to cumulative voting, however, the increase of the points on an option as a result of expressing stronger support is not linear but quadratic~\citep{posner2015voting}. This means that expressing a strong preference would be more costly~\citep{goodman2021will} (see Fig.~\ref{fig:com}). 

Despite their potential benefits, quadratic and cumulative voting are not widely adopted in most voting platforms. Most voting platforms tend to use more straightforward voting methods such as plurality or majority voting~\citep{cooper2012comparison,Pournaras2020}. These methods are simpler to understand and administer, but they do not provide the same level of flexibility as cumulative and quadratic voting. We successfully implemented cumulative and quadratic voting on the Stanford Participatory Budgeting platform\footnote{Available at \url{https://github.com/DISC-Systems-Lab/SPB} (last access: July 2023).}~\citep{PB-Stanford2015}, a popular open-source platform used for several campaigns in North America, and recently in Europe (i.e. Aarau city in Switzerland). The effective implementation of cumulative and quadratic voting relies not only on the underlying voting method, but also on the design and usability of the voter interface. 

\section{Case Study: Cumulative and Quadratic Voting}

Cumulative and quadratic voting are a group of flexible multi-option preferential voting methods that enable participants to distribute a number of points across multiple options based on the intensity of their preferences. This flexibility allows voters to express their support for a range of options, rather than being limited to a single choice, as is highlighted in Fig.~\ref{fig:com}. As a result, the voters can prioritize their preferred options, putting more weight on their top choices while still compromising and providing support to other options they find acceptable. The feature that distinguishes cumulative and quadratic voting from other voting approaches is its ability to allow voters to make complex trade-offs among the options they support, i.e. more points on the most preferred option exhaust the available points left for other options and vice versa. This design enables cumulative and quadratic voting to encode and capture a richer amount of information regarding voter preferences.

Moreover, cumulative voting provides a voice to minorities by empowering them to concentrate their voting power on specific choices that are important to them. In other words, they can allocate more weight to a topic that holds significant importance and sensitivity for them. This mechanism ensures that their preferences are not marginalized by the decisions of the majority, thereby providing a more inclusive decision-making process~\citep{casella2023minority}.

\section{Voting Interface and Voter Experience}

The Stanford Participatory Budgeting interface supported until recently by default k-approval voting, k-ranking voting, knapsack voting and pairwise comparison voting. Implementing cumulative and quadratic methods required an additional user interface (UI) to assign a number of tokens, keep track of tokens assigned, remove tokens, and redistribute them to other projects. To have an intuitive voter experience, the designed UI should simplify the process of point allocation, tracking, and redistribution. The following provides some requirements for implementing such a UI:
\begin{itemize}
    \item \textbf{Point allocation:} Providing voter-friendly interface elements (i.e., buttons or sliders) that allow a voter to set the number of tokens they choose to assign to each option. The UI should reflect the remaining tokens available and update dynamically as voters allocate their tokens.
    
    \item \textbf{Points tracking:} Display a visual representation or numerical count of the tokens assigned to each option. This could be shown as a progress bar, a numerical value, or other intuitive visualization that allows voters to easily see the distribution of their tokens.

    \item \textbf{Cost:} Voters are able to make trade-offs between their prefered projects and the token budget exhaustion via a fine-grained control.  
    
    \item \textbf{Removing and redistributing tokens:} Include interactive controls that enable voters to remove tokens from an option and add them elsewhere to adjust their allocations in real-time. The UI should reflect the changes made and update points tracking accordingly.
\end{itemize}

The platform should ensure that voters do not exceed the total number of tokens available or encounter any errors while assigning or redistributing tokens. It requires clear feedback messages and notifications to guide voters and prevent any unintended mistakes. Testing and voter feedback can also be valuable in refining the UI design to ensure a smooth and efficient voter experience. In this context, we conducted an experiment to compare voter satisfaction regarding the UI flexibility in keeping track of tokens assigned to the options. The experiment aimed to assess how different UI designs or variations impact user experience and the effectiveness of cumulative and quadratic voting. For simplicity, we opted to focus on the cumulative voting method. As cumulative voting shares a number of similarities with quadratic voting, the findings can be generalized to quadratic voting and other distributional methods with the support of some tailored additional explanations. 

\begin{figure}[!htb]
\centering
\begin{subfigure}{.49\textwidth}
  \centering
  \includegraphics[width=0.99\linewidth]{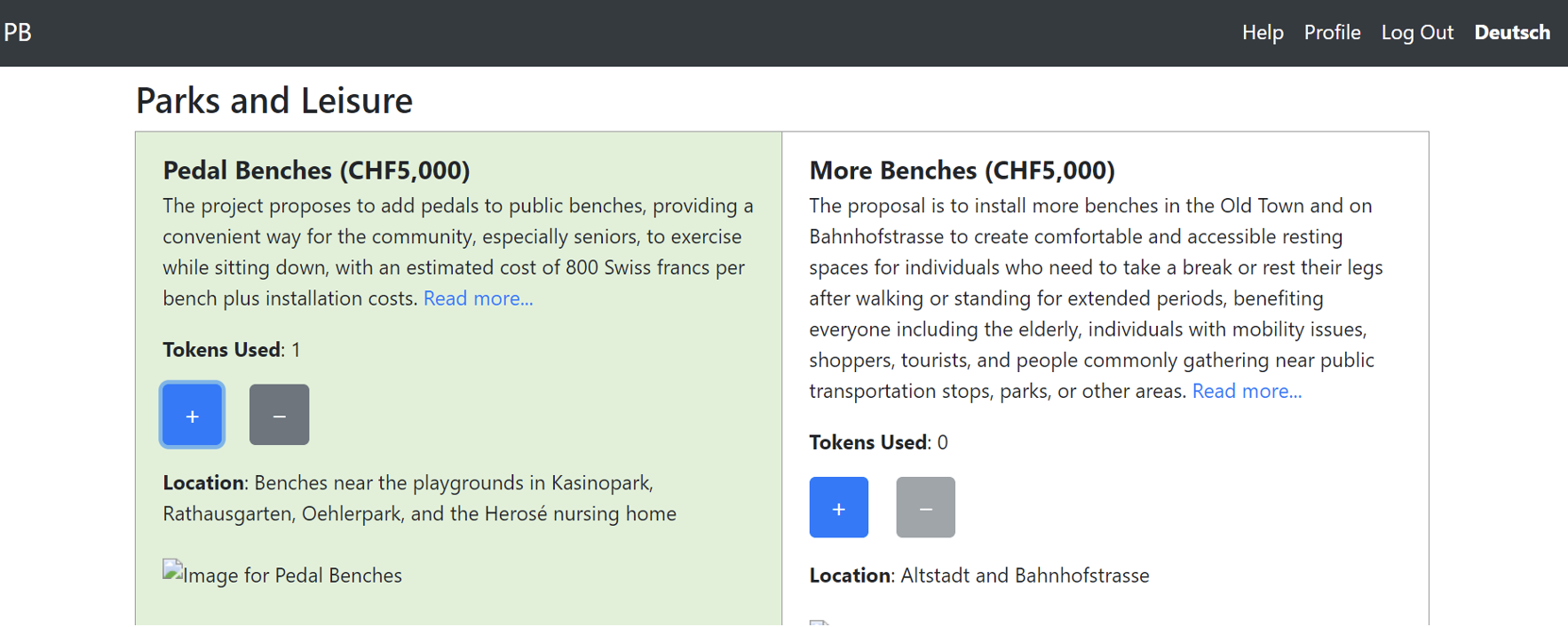}
  \caption{A simple UI.}
  \label{fig:f2}
\end{subfigure}
\begin{subfigure}{.49\textwidth}
  \centering
  \includegraphics[width=.82\linewidth]{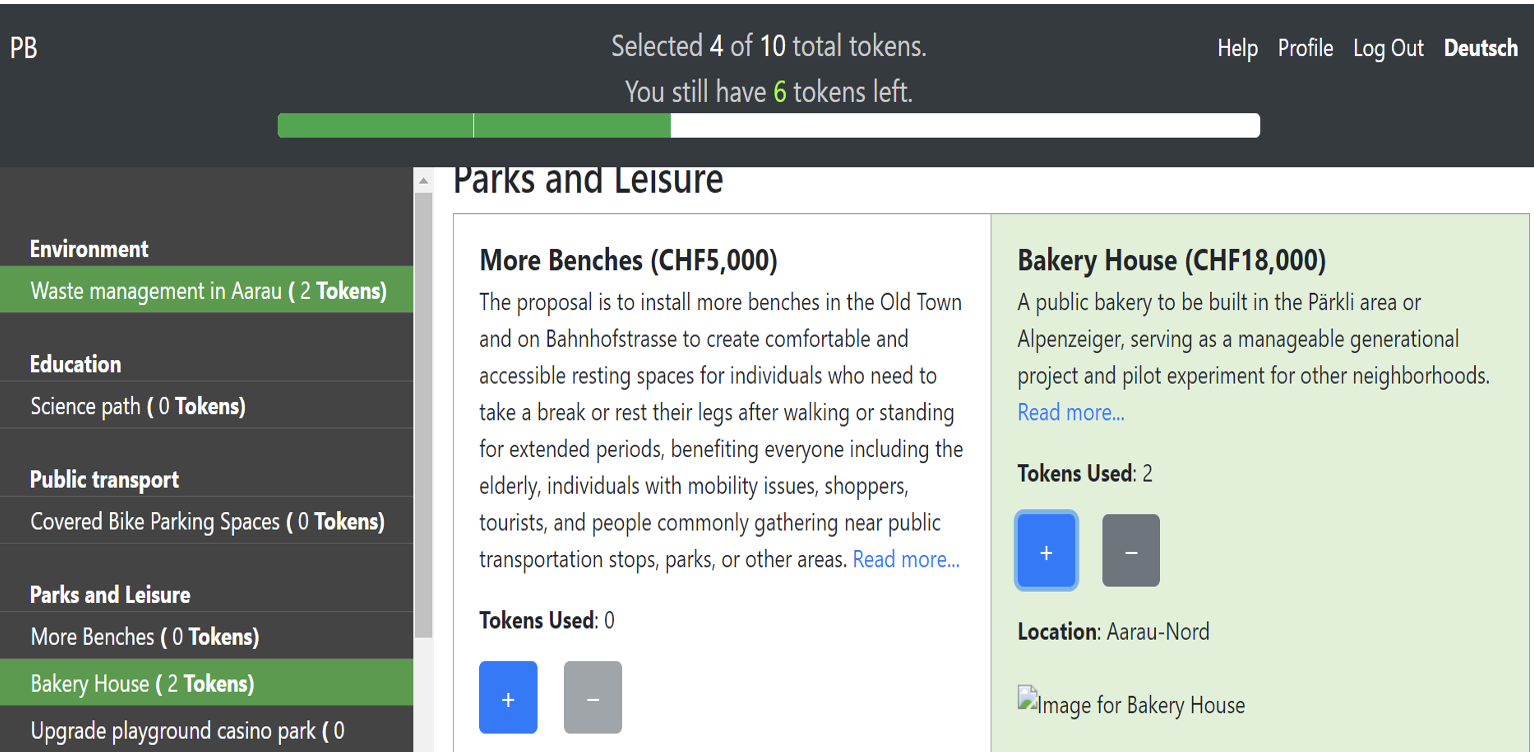}
  \caption{An augmented UI.}
  \label{fig:f1}
\end{subfigure}%
\caption{UI comparison for cumulative voting. The point and side bar show the number assigned and remaining tokens.}
\label{fig:test}
\end{figure}

During the experiment, participants were presented with different versions of the UI, each offering varying degrees of flexibility and ease of use in terms of points allocation and tracking, see Fig.~\ref{fig:f2} and \ref{fig:f1}. The side and point bars are borrowed graphical elements from the Knapsack voting of Stanford Participatory Budgeting

\section{Empirical Study}

\subsection{Design}

A study was conducted using Qualtrics and sent to 90 Media and Communication students from the University of Fribourg, Switzerland. In total, 27 students responded. We asked the survey respondents to vote in two voting scenarios using the Stanford Participatory Budgeting platform (k-ranking\footnote{This is implemented using a k-approval interface combined with a ranking step} and cumulative vote), with follow-up questions assessing their preference for the voting method and UI design.

\subsection{Results}

Table~\ref{tab:results} provides an outline of the results of the study. We began by asking respondents to indicate, on a scale of one to ten, how well ranked voting and cumulative voting allowed them to express their preferences. The mean response for ranked voting was 6.06 and cumulative voting was 7.63, suggesting respondents felt that cumulative voting was a more effective method to express their preferences. A t-test was conducted to compare the mean scores for ranked voting and cumulative voting. The test had a p-value of 0.15, demonstrating significance at 95\% confidence. 

In the following question, 75\% of respondents suggested that cumulative voting was their preferred voting method, 19\% suggested that ranked voting was their preferred method and 6\% were unsure. We then asked a semantic differential question, in which respondents rate their preference between a challenging method that was more accurate in expressing preferences and a simple method that was less accurate in expressing preferences. The normalized mean value in the range $[0,1]$ presented was 0.687 (where a value of below .5 represents favorability towards a complex but more accurate voting method and a value of greater than .5 represents favorability towards a simple but less accurate voting method). In this sense, the value of 0.687 highlights that respondents where favourable towards a simple but less accurate voting method. For these results a t-test was conducted and had a p-value of <.0001 demonstrating significance at 95\% confidence. 

\begin{table}[!htb]
\centering
\begin{tabularx}{\textwidth}{cYc}
\toprule
\textbf{Question Number} & \textbf{Question} & \textbf{Results} \\
\midrule
1 & Please indicate how well you were able to express your preferences with ranked voting (0-10) & 6.06 (mean) \\
\addlinespace
2 & Please indicate how well you were able to express your preferences with cumulative voting (0-10) & 7.63 (mean) \\
\addlinespace
3 & Indicate your preferred voting method & Ranked Voting = 19\% \\
   & & Cumulative Voting = 75\% \\
   & & Unsure = 6\% \\
\addlinespace
4 & Please indicate your preference between a complex but accurate (0) and simple but less accurate voting method (1) & 0.687 (mean) \\
\addlinespace
\multirow{4}{*}{} 5 & Which is the most helpful when voting? & No Additional Interface = 0\% \\
   & & Top Bar = 26\% \\
   & & Side Bar = 6\% \\
   & & Top and Side Bar = 68\% \\
\bottomrule
\end{tabularx}
\caption{Overview of the study results.}
\label{tab:results}
\end{table}

\begin{figure}
  \centering
\includegraphics[width=0.9\textwidth]{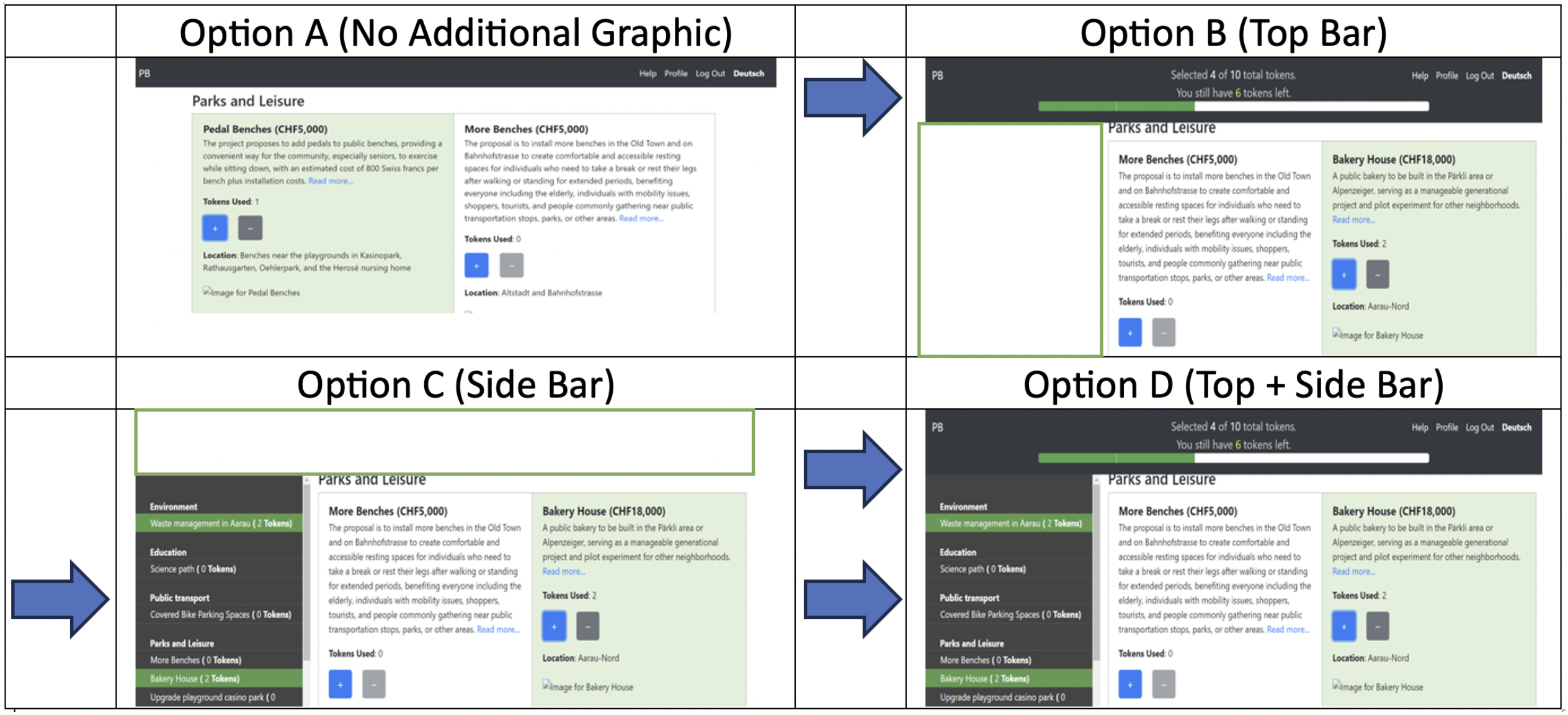}
  \caption{User interface question.}
  \label{fig:ui}
\end{figure}

Finally, to assess the changes that were made to the user interface to convey information from the more complex voting methods, we asked respondents to select their preferred user interface, between the four layouts highlighted in Fig.~\ref{fig:ui} (participants were exposed to Option C in k-ranking and Option D in cumulative voting in the voting exercise). 26\% selected Option B (Top Bar), 6\% selected Option C and 68\% selected Option D (Top + Side Bar). Option A (No Additional Graphics) received no support.

\section{Conclusion and Future Work}

In this study, we sought to address challenges with cumulative and quadratic voting on digital voting platforms, as well as assess the impact of the voter interface representing information from such voting methods. To do so, we implemented cumulative and quadratic voting on the digital voting platform Stanford Participatory Budgeting, and adapted the user interface. Through an experiment we find that cumulative voting, although often overlooked due to its perceived complexity, emerged as a preferred method over ranked voting according to the empirical findings. Interestingly, while participants regarded cumulative voting as superior, they also expressed preference for simple but less accurate voting methods. 

These findings suggest that even with its complex nature, the effective implementation of cumulative and quadratic voting is feasible. Participants' preference towards simplicity in voting processes indicates the potential benefits of platform-level solutions aimed at simplifying cumulative voting. Additionally, our research highlights that tailored interface features could enhance voter satisfaction.

By analyzing user feedback and preferences, we have presented a case for the integration of cumulative and quadratic voting in digital platforms. This study underscores the importance of voter interface design in shaping voter interaction and experience, potentially leading to broader adoption of more expressive voting methods such as cumulative and quadratic voting. In doing so, the findings from this paper have implications for fairness and inclusion, as cumulative and quadratic voting can encourage fair and inclusive results~\citep{ricnckevivcs2013equality, casella2023minority}.  

We will build on the findings from the paper through research conducted in a field test in Aarau, Switzerland. The field test has used the cumulative voting on Stanford PB platform, with the adapted user-interface changes outlined within this paper, in a real-world participatory budgeting vote. This should grant further information as to the impact of distributional voting methods on participatory budgeting and strengthen the findings presented within this paper. Future research will also study and compare the different interface interventions required between cumulative and quadratic voting.

\section{Acknowledgements}

We would like to acknowledge the other members of the Stanford Crowdsourced Democracy Team for the development of the Stanford Participatory Budgeting platform and their support to this project, in particular Ashish Goel, Sukolsak Sakshuwong and Sahasrajit Sarmasarkar. We would also like to thank Jasmin Odermatt, Lea Good, Sebrina Pedrossi and Joshua Yang for providing their insights on improvements to the cumulative voting interface. This work is supported by a UKRI Future Leaders Fellowship (MR\-/W009560\-/1): `\emph{Digitally Assisted Collective Governance of Smart City Commons--ARTIO}' 
and
by the SNF NRP77 `Digital Transformation' project "Digital Democracy: Innovations in Decision-making Processes", \#407740\_187249. 

%\printbibliography

\bibliography{main}

\begin{thebibliography}{14}
\providecommand{\natexlab}[1]{#1}
\providecommand{\url}[1]{\texttt{#1}}
\expandafter\ifx\csname urlstyle\endcsname\relax
  \providecommand{\doi}[1]{doi: #1}\else
  \providecommand{\doi}{doi: \begingroup \urlstyle{rm}\Url}\fi

\bibitem[Pournaras(2020)]{Pournaras2020}
Evangelos Pournaras.
\newblock Proof of witness presence: Blockchain consensus for augmented
  democracy in smart cities.
\newblock \emph{Journal of Parallel and Distributed Computing}, 145:\penalty0
  160--175, 2020.

\bibitem[Helbing et~al.(2023)Helbing, Mahajan, Fricker, Musso, Hausladen,
  Carissimo, Carpentras, Stockinger, Sanchez-Vaquerizo, Yang,
  et~al.]{Helbing2023}
Dirk Helbing, Sachit Mahajan, Regula~H{\"a}nggli Fricker, Andrea Musso,
  Carina~I Hausladen, Cesare Carissimo, Dino Carpentras, Elisabeth Stockinger,
  Javier~Argota Sanchez-Vaquerizo, Joshua~C Yang, et~al.
\newblock Democracy by design: Perspectives for digitally assisted,
  participatory upgrades of society.
\newblock \emph{Journal of Computational Science}, 71:\penalty0 102061, 2023.

\bibitem[Junior et~al.(2022)Junior, Ferreira, et~al.]{junior2022informative}
Eli~Borges Junior, Bruno~Madureira Ferreira, et~al.
\newblock Informative architectures and citizen participation: a comparative
  study between the digital platforms decidim and rousseauv.
\newblock \emph{Liinc em Revista}, 18\penalty0 (2):\penalty0 e6058--e6058,
  2022.

\bibitem[Wellings et~al.(2023)Wellings, Majumdar, Haenggli~Fricker, and
  Pournaras]{wellings2023improving}
Thomas~Samuel Wellings, Sirjoni Majumdar, Regula Haenggli~Fricker, and
  Evangelos Pournaras.
\newblock Improving city life via legitimate and participatory policy-making: A
  data-driven approach in switzerland.
\newblock In \emph{Proceedings of the 24th Annual International Conference on
  Digital Government Research}, pages 23--35, 2023.

\bibitem[Benade et~al.(2017)Benade, Nath, Procaccia, and
  Shah]{benade2017preference}
G.~Benade, S.~Nath, A.~Procaccia, and N.~Shah.
\newblock Preference elicitation for participatory budgeting.
\newblock In \emph{Proceedings of the 31st AAAI Conference on Artificial
  Intelligence (AAAI-17)}, pages 376--382, 2017.

\bibitem[Goel et~al.(2019)Goel, Krishnaswamy, Sakshuwong, and
  Aitamurto]{goel2019knapsack}
Ashish Goel, Anilesh~K Krishnaswamy, Sukolsak Sakshuwong, and Tanja Aitamurto.
\newblock Knapsack voting for participatory budgeting.
\newblock \emph{ACM Transactions on Economics and Computation (TEAC)},
  7\penalty0 (2):\penalty0 1--27, 2019.

\bibitem[Kato et~al.(2021)Kato, Asa, and Owa]{kato2021positionality}
Takeshi Kato, Yasuhiro Asa, and Misa Owa.
\newblock Positionality-weighted aggregation methods for cumulative voting.
\newblock \emph{Int'l J. Soc. Sci. Stud.}, 9:\penalty0 79, 2021.

\bibitem[Lalley and Weyl(2014)]{lalley2014nash}
Steven~P Lalley and E~Glen Weyl.
\newblock Nash equilbria for quadratic voting.
\newblock \emph{arXiv preprint arXiv:1409.0264}, 2014.

\bibitem[Ri{\c{n}}{\c{k}}evi{\v{c}}s and
  Torkar(2013)]{ricnckevivcs2013equality}
K~Ri{\c{n}}{\c{k}}evi{\v{c}}s and Richard Torkar.
\newblock Equality in cumulative voting: A systematic review with an
  improvement proposal.
\newblock \emph{Information and Software Technology}, 55\penalty0 (2):\penalty0
  267--287, 2013.

\bibitem[Casella et~al.(2023)Casella, Guo, and Jiang]{casella2023minority}
Alessandra Casella, Jeffrey Da-Ren Guo, and Michelle Jiang.
\newblock Minority turnout and representation under cumulative voting. an
  experiment.
\newblock \emph{Games and Economic Behavior}, 2023.

\bibitem[Posner and Weyl(2015)]{posner2015voting}
Eric~A Posner and E~Glen Weyl.
\newblock Voting squared: Quadratic voting in democratic politics.
\newblock \emph{Vand. L. Rev.}, 68:\penalty0 441, 2015.

\bibitem[Goodman and Porter(2021)]{goodman2021will}
John~C Goodman and Philip~K Porter.
\newblock Will quadratic voting produce optimal public policy?
\newblock \emph{Public Choice}, 186\penalty0 (1-2):\penalty0 141--148, 2021.

\bibitem[Cooper and Zillante(2012)]{cooper2012comparison}
Duane Cooper and Arthur Zillante.
\newblock A comparison of cumulative voting and generalized plurality voting.
\newblock \emph{Public Choice}, 150:\penalty0 363--383, 2012.

\bibitem[{PB-Stanford}(2015)]{PB-Stanford2015}
{PB-Stanford}.
\newblock Stanford participatory budgeting platform, 2015.
\newblock URL \url{https://pbstanford.org/}.

\end{thebibliography}

\end{document}